\documentclass{article}
\usepackage{amsmath,epsfig,amssymb}
\usepackage{graphicx}
\usepackage{eufrak,mathrsfs}
\usepackage[mac]{inputenc}
\usepackage{array} 
\usepackage{upgreek,bm}

\def\x{{\boldsymbol x}}
\def\a{{\boldsymbol a}}
\def\r{{\boldsymbol r}}
\def\rr{{\boldsymbol {r'}}}
\def\b{{\boldsymbol b}}
\def\k{{\boldsymbol k}}

\def\m{{\boldsymbol m}}

\newcommand{\interp}{\mathcal I}

\newcommand{\mbb}{\mathbf}

\newcommand{\inner}[1]{\left\langle #1\right\rangle}

\newcommand{\h}{\widehat}

\newcommand{\btau}{\boldsymbol{\tau}}
\newcommand{\bbeta}{\boldmath{\beta}}

\long\def\symbolfootnote[#1]#2{\begingroup%
\def\thefootnote{\fnsymbol{footnote}}\footnote[#1]{#2}\endgroup}

\newtheorem{theorem}{Theorem}[section]

\newtheorem{proposition}[theorem]{Proposition}

\newtheorem{definition}[theorem]{Definition}

\title{Fast adaptive elliptical filtering using box splines}

\author{Kunal N. Chaudhury$^{\star}$, Arrate M. Barrutia$^{\diamond}$, Michael Unser$^{\star}$\\
$\star$ Biomedical Imaging Group (BIG),\\
 \'Ecole Polytechnique Fédérale de Lausanne (EPFL), Switzerland. \\ 
$\star$ Center for Applied Medical Research (CIMA), \\
University of Navarra, Pamplona, Spain}

\date{}

\begin{document}

\maketitle

\begin{abstract}
   	We demonstrate that it is possible to filter an image with an elliptic window of varying size, elongation and orientation with a fixed computational cost per pixel. Our method involves the application of a suitable global pre-integrator followed by a pointwise-adaptive localization mesh. We present the basic theory for the $1$D case using a B-spline formalism and then appropriately extend it to $2$D using \textit{radially-uniform box splines}. The size and ellipticity of these radially-uniform box splines is adaptively controlled. Moreover, they converge to Gaussians as the order increases. Finally, we present a fast and practical directional filtering algorithm that has the capability of adapting to the local image features.     
\end{abstract}

\section{Introduction}
The most common smoothing operator is the Gaussian filter. For that reason, it is of practical interest to design efficient directional filtering strategies based on this type of filter. Fast recursive solutions for space-invariant Gaussian-like filtering have been developed \cite{Smeulders} but the space-variant ones are subject of current research. To date, two algorithms have been proposed:  one that works in $1$D and uses B-splines for fast computations of the Continuous Wavelet Transform \cite{Munoz}, and the other proposed in Computer Graphics that essentially does a rectangular smoothing using repeated integration \cite{Heckbert}.  
		
In this paper, we propose a general technique for $N$-directional adaptive filtering using box spline formalism and discuss its implementation for the special four-directional case. Although the derivation is rather involved, the final solution is quite simple (convolution-like with an adaptive mesh) and rather easy to implement (cf. Eqn. \eqref{final_eqn}). Also, the algorithm has a constant computational cost per pixel as the support of the adaptive mesh is independent of the pointwise-adaptive scale-vector. 
	
The paper is organized as follows. In $\S$ 2, we revisit B-splines and certain associated linear operators. In $\S$ 3, we describe an efficient B-spline based adaptive filtering technique, initially proposed in \cite{Munoz}. We then introduce the family of radially-uniform box splines and some related linear operators in $\S$ 4 and $\S$ 5, respectively. Finally, in $\S$ 6, we propose the adaptive directional filtering strategy followed by the description of a fast algorithm in $\S$ 7.

\section{$1$D Linear Operators and B-splines}
We first introduce two linear, shift-invariant operators, initially defined for real-valued functions $f(x)$ and then appropriately applied to real-valued sequences\footnote{By sequences we will mean functions defined on the Cartesian lattice $\mbb{Z}^d$, where $d$ is the dimensionality of the signal.} $g[k]$.   
\begin{definition}
The finite-difference (FD) operator $\Delta^{n}_{a}$ of order $n \in \mathbb{Z}_+$ and scale $a \in  \mbb{R}_+$ is specified as
\begin{align}
\label{FD}
\Delta^{n}_{a} f(x)=\sum_{k=0}^n  d^n_a[k] f(x-ak) 
\end{align}
where $d^n_a[k]=a^{-n} (-1)^k \binom{n}{k}$, for $0 \leq k \leq n$ and $0,$ else.
\end{definition}
In the Fourier domain, we have: $\widehat{\Delta^{n}_{a} f}(\omega)=\ \h {\Delta}_a^{n}(e^{ja\omega}) \h {f}(\omega)$, with $\h {\Delta}_a^{n}(e^{ja\omega})=\sum_k d^n_a[k] e^{-ja\omega k}=a^{-n} (1-e^{-ja \omega })^n$ interpreted as the $(2\pi/a)$-periodic frequency response of the FD filter $d^n_a[k]$. 

When acting on sequences $g[k]$ , the FD filter acts as a discrete convolution operator. For integer $a$, we have $\Delta^{n}_{a}g[m]=\sum_{k=0}^{n} d^{n}_{a}[k] g[m-ak]$; for non-integer $a$, some form of interpolation is necessary as $g$ is not defined for non-integer arguments. 
\begin{definition}
The running-sum (RS) operator $\Delta_b^{-1}$ of scale $b \in \mbb{R}_+$ is given by
\begin{align}
\label{rs}
\Delta_b^{-1} f(x)=b\sum_{k=0}^{\infty} f(x-bk).
\end{align}
\end{definition}
For integer $b$, the RS operator applied to sequences $g[k]$ corresponds to the digital filter $u_b[k]=b$, for $k \in b\mbb{N}_0$ and $0,$ otherwise, with the correspondence: $\Delta_b^{-1} g[n]=\sum_{k \in \mbb{Z}} u_b[k]g[n-k]$. Note that $y=u_b \ast g$ can be implemented very efficiently using the recursive equation $y[m]=y[m-b]+bg[m]$, with appropriate boundary conditions \cite{Munoz}. The $n$-fold composition of the above RS operator will be denoted by $\Delta_b^{-n}$ with frequency response $\h {\Delta}_b^{-n}(e^{jb\omega})=b^n(1-e^{-jb\omega})^{-n}$; for integer $b$, the corresponding filter $u_b^{n}[k]$ is specified by $\sum_{k \in \mbb{Z}} u_b^{n}[k] e^{-j\omega  k}=b^n(1-e^{-jb\omega})^{-n}$. 

Finally, we introduce the family of symmetric B-splines \cite{spline} that are closely related to the above linear operators; it is the following Fourier domain definition that makes the link apparent. 
\begin{definition}
The symmetric B-spline $\beta_a^{n}(x)$ of degree $n \in \mathbb{N}_0$ and of scale $a \in \mbb{R}_+$ is specified by the Fourier transform 
\begin{align}
\label{definition}
\h{\beta}^{n}_a(\omega)=\frac{1}{a^{n+1}} \left(\frac{e^{ja\omega/2}-e^{-ja\omega/2}}{j\omega}\right)^{n+1}
\end{align}
\end{definition}
Specifically, B-splines of arbitrary scales can be expressed in terms of integer-scaled B-splines:  
\begin{align}
\label{link}
\beta_{a}^{n}(x)=\left(\Delta_{a}^{n+1} \circ \Delta_{1}^{-(n+1)}\right)\beta_{1}^{n}(x+\tau)
\end{align}
with $\tau=(a-1)(n+1)/2$, using the FD and RS operators \cite{Munoz}. It is also worth mentioning that $\beta^0_a(x)=1/a,$ for $x \in (-a/2,a/2]$ and $0$ else; it is the piecewise-constant function $\mbox{rect}(x/a)$. In the sequel, we simply denote it by $\beta_a(x)$.

\section{Scale-Adaptive Filtering}
	
In this section, we revisit our $1$D space-variant filter \cite{Munoz} based on the projections $s[m]=\inner{f(x),\beta_a^{n_2}(x-m)}, m \in \mbb{Z},$ of a continuous signal model $f(x)$ of the discrete signal with scaled B-splines $\beta_{a}^{n_{2}}$. The scale $a$ controls the degree of smoothing applied around each sample.

	Specifically, given the signal samples $f[k]$, we consider the following B-spline model $f(x)=\sum_{k \in \mbb{Z}} c[k] \beta^{n_1}(x-k)$ for the continuum, with the interpolation constraint $f(x){|}_{x=k}$  $=f[k]$, $k \in \mbb{Z}$. The expansion coefficients are then obtained by the digital filtering $c=f \ast {(b^{n_1})}^{-1}$, where ${(b^{n_1})}^{-1}$ is the convolution inverse of the B-spline interpolation filter $b^{n_1}[k]=\beta^{n_1}(k), k \in \mbb{Z}$ \cite{spline}. It then turns out that the projections $s[m]$ can be efficiently realized using \eqref{link}:
\begin{proposition}
\label{prop}
The B-spline projection $s[m]$ can be computed in two steps:
\begin{enumerate}
\item \textbf{Non-Adaptive Step:} The running sum filter $u_1^{(n_2+1)}$ is applied to all the filtered sequence $c[k]$ to get the integrated sequence 
\begin{equation*}
g[m]=\sum_{k \in \mbb{Z}}  u_1^{(n_2+1)}[k]  c[m-k].
\end{equation*}
Then, the continuous domain pre-integrated signal can be written as $F(x)=\sum_{k \in \mbb{Z}} g[k] \beta^{n_1+n_2+1}(x+\tau-k)$.
\item \textbf{Adaptive Step:}
At each sample position, $m\in\mathbb{Z}$, and corresponding to a specific scale $a$, the integrated sequence $g$ is then filtered using the FIR localization mask $w[k]=\Delta_a^{n_2+1}\beta^{n_1+n_2+1}(k+\tau),k \in \mbb{Z}$ to get the local projection $\label{filtered} s[m]=\Delta_a^{n_2+1}F(m)=\sum_{k \in \mbb{Z}}  g[k] w[m-k]$.
\end{enumerate}
\end{proposition}

\section{Radially-Uniform Box Splines}
We now extend these ideas to $2$D, where the additional feature of directionality needs to be addressed. Particularly, we devise the following tensor product between a $1$D smoothing operator and the identity convolution operator $\delta: \psi \mapsto \psi(0)$, both operating along orthogonal directions.
\begin{definition}

The generating kernel $\varphi_a$ of scale $a \in \mbb{R}_+$ is defined as  
\begin{align}
\label{generator}
\varphi_a(\x)=\beta_a(x_1)\delta(x_2), \ \x=(x_1, x_2) \in \mathbb{R}^2 
\end{align}
\end{definition}

The rotated versions $\varphi_{a,\theta}(\x)$ of the generating kernel are then obtained as
\begin{align*}
\varphi_{a,\theta}(\x)=\boldsymbol{R}_{\theta}\varphi_a(\x)=\beta_a (\r_{\theta}^T \x) \delta(\rr_{\theta}^T \x)  
\end{align*}
where $\boldsymbol{R}_{\theta}$ is the rotation operator  defined as $\boldsymbol{R_{\theta}}f(\x)=f(R^T_{\theta}\x)$ via the rotation matrix
\begin{align*}
R_{\theta}= \left(\begin{array}{cc}\cos{\theta} & -\sin{\theta} \\ \sin{\theta} & \cos{\theta}\end{array}\right)=\Big(\begin{array}{c}\boldsymbol{{r_{\theta}}} \  \   \boldsymbol{{r'_{\theta}}}  \end{array}\Big)  
\end{align*}
The most elementary box spline, the $\mbox{rect}(\x)$ function, can be generated by convolving two quadrature tensor B-splines; specifically, $\mbox{rect}(\x)=\varphi_{1,0}(\x) \ast \varphi_{1,\pi/2}(\x)$ as per the above formalism. We generalize this idea to construct a family of box splines, which we call the \textit{radially-uniform box splines}. 
\begin{definition}
The radially-uniform box spline $\bbeta^N_{\boldsymbol{a}}$ is defined to be
\begin{align}
\label{def}
\bbeta^N_{\boldsymbol{a}} (\x)= (\varphi_{a_1,\theta_1}\ast \cdots \ast \varphi_{a_N,\theta_N})(\x)
\end{align}
where $N \in \mbb{N}, N \geq 2$ is the directional order, $\a=(a_1,\ldots,a_N) \in \mbb{R}^N_+$ is the scale vector, and $\theta_k=(k-1)\pi/N, 1\leq k \leq  N$ are the rotation angles.
\end{definition}
Intuitively this means that we can construct $2$D box splines by convolving arbitrary number of tensor B-splines, rescaled by arbitrary amounts but uniformly distributed radially. In retrospect, $\mbox{rect}(\x)=\bbeta^2_{(1,1)}(\x)$. 

Importantly, the covariance (moment) of  these smoothing kernels can be arbitrarily controlled by suitably choosing the order $N$ and the pairs $\{(a_k,\theta_k)\}_{k=1}^N$. The remarkable fact is that as the order increases, these box splines become more Gaussian-like. In particular, we have the following result:
\begin{theorem}
\label{convergence}
Let $\{\bbeta^N_{\a(N)}\}_{N \geq 2}$ be a sequence of box splines corresponding to the the scale-vector sequence $\{\a(N)\}_{N \geq 2}$ with components given by $a_k(N)= \sigma \sqrt{24/N}, 1\leq k \leq N$. Then we have the following convergence
\begin{align}
\label{conv}
 \lim_{N \rightarrow \infty} \bbeta^N_{\a(N)}(\x)=\frac{1}{2\pi \sigma^2} \exp\left(-\frac{||\x||^2}{2\sigma^2} \right).
\end{align}
\end{theorem}

Yet another form of convergence is achievable based on the iterated convolution of the box spline $\bbeta^N_{\a}$, of a specific order $N$, with itself. Based on the Central Limit Theorem, existence of a sequence of iterated box splines that converges to a Gaussian can also be demonstrated. 
\section{$2$D Linear Operators}

 We now extend the definitions of the FD \eqref{FD} and the RS \eqref{rs} operator to $2$D; the key aspect is to preserve \eqref{link} in relation to the radially-uniform box-splines. 
\begin{definition}
The FD operator $\Delta_{a,\theta}$ of scale $a \in \mbb{R}_+$ operating in the direction $\theta \in [0,\pi)$ is specified by
\begin{align*}
\Delta_{a,\theta}f(\x)=\frac{1}{a}\Big(f(\x)-f(\x-a\r_{\theta})\Big).
\end{align*}
\end{definition} 

  The $N$-directional FD operator $\Delta^N_{\a}$ corresponding to the scale-vector $\a=(a_1,\ldots,a_N)$ is then defined as the composition $\Delta^N_{\a}=\Delta_{a_1,\theta_1} \circ \cdots \circ \Delta_{a_N,\theta_N}$, where $\{\Delta_{a_k,\theta_k}\}_{k=1}^N$ are FD operators of scale $a_k \in \mbb{R}_+$ operating in the direction $\theta_k=(k-1)\pi/N$, repectively. 
\begin{definition}
The RS operator $\Delta^{-1}_{b,\theta}$ of scale $b \in \mbb{R}_+$ operating in the direction $\theta \in [0,\pi)$ is given by
\begin{align}
\label{RS1}
\Delta^{-1}_{b,\theta}f(\x)=b\sum_{k =0}^{\infty} f(\x-kb\r_{\theta}).
\end{align}
\end{definition}

The $N$-directional RS operator $\Delta^{-N}_{\b}$ corresponding to the scale-vector $\b=(b_1,\ldots,b_N)$ is then defined as $\Delta^{-N}_{\b}=\Delta^{-1}_{b_1,\theta_1} \circ \cdots \circ \Delta^{-1}_{b_N,\theta_N}$, where $\{\Delta^{-1}_{b_k,\theta_k}\}_{k=1}^N$ are RS operators of scale $b_k$ operating in the direction $\theta_k=(k-1)\pi/N$, respectively. 

 Importantly, based on \eqref{RS1}, the application of $\Delta^{-N}_{\b}$  on a discrete sequence $g[\k]$ is then (non-uniquely) given by
\begin{align} 
\label{def_rs}
\Delta^{-N}_{\b} g[\m]=b_1\sum_{k_1=0}^{\infty} \cdots b_N \sum_{k_N=0}^{\infty} g_{\text{int}}\Big(\m-\sum_{j=1}^N k_jb_j\r_{\theta_j}\Big)
\end{align}
where $g_{\text{int}}(\x)$ is some form of interpolation of the discrete sequence $g[\k]$. The good news is that for a specific choice of $N$ and $\b$, the  operator $\Delta^{-N}_{\b}$ admits a convolution kernel (filter) and \eqref{def_rs} then has a unique form; this motivates the following definition and subsequent simplification.

Let $0 \leq \theta < \pi$ and $b \in \mbb{R}_+$ be such that $b\cos \theta,b\sin \theta \in \mbb{Z}$. Then the RS filter $u_{b,\theta}[\k]$ of scale $b$ and acting in the direction $\theta$ is defined as
\begin{align}
\label{RS_filter}
u_{b,\theta}[\k]=\begin{cases}  b, &  \text{for \ $\k \in (b\cos\theta,b\sin\theta)\mbb{N}_0$} \\
0, &\text{else}
\end{cases}
\end{align}
The $N$-directional RS filter $u^N_\b[\k]$ corresponding to the integration scale-vector $\b=(b_1,\ldots,b_N)$ is then defined as the $2$D convolution 
$u^N_{\b}=u_{b_1,\theta_1} \ast \cdots \ast u_{b_N,\theta_N}$ of the RS filters $\{u_{b_k,\theta_k}\}_{k=1}^N$  of scale $b_k$ and acting in the direction $\theta_k$; provided all the constituent filters are well-defined. Then \eqref{def_rs} is uniquely given by $\Delta^{-N}_{\b} g[\m]=\sum_{\k \in \mbb{Z}^2} u^N_{\b}[\k] g[\m-\k]$.

Importantly, as in the $1$D case, it turns out that $\bbeta^N_{\boldsymbol{a}}$ can be factorized using the above linear operators as follows
\begin{align}
\label{factor}
\bbeta^N_{\boldsymbol{a}}(\x) =\left(\Delta^N_{\a} \circ \Delta^{-N}_{\b} \right) \bbeta^N_{\boldsymbol{b}}(\x+\btau)
\end{align}
where $\btau=0.5\Big(\sum_{k=1}^N (a_k-b_k)\cos \theta_k, \sum_{k=1}^N (a_k-b_k)\sin \theta_k\Big)$ is the shift-vector. The role of the scale-vector $\b$ in \eqref{factor} is to tie the RS operator $\Delta^{-N}_{\b}$ to the underlying lattice $\mbb{Z}^2,$ and will be used to advantage in the sequel.  

\section{Space-Variant Adaptive Filtering}
		Our goal is to adaptively filter the signal $f(\x)$, obtained by interpolating the $2$D signal samples $f[\k]$, with appropriately elongated and orientated box splines $\bbeta^N_{\a}(\x)$ by suitably adjusting the scale-vector $\a$ at the different spatial locations. In other words, given some pre-assigned scale-vector map $\a: \mbb{Z}^2 \rightarrow \mbb{R}^N_+$, we want to compute the projections $s[\m]=\langle f,\bbeta_{\a(\m)}^N(\cdot-\m) \rangle$, at spatial locations $\m \in \mbb{Z}^2$. 

	First, as in the $1$D case, we consider the continuous representation $f(\x)=\sum c[\boldsymbol{k}] $ $\phi(\x-\boldsymbol{k})$, where $\phi(\x)$ is a $2$D interpolating function. The expansion coefficients are given by $c=f \ast {(b^{n})}^{-1}$, where $(b^n)^{-1}$ is the convolution inverse of the interpolation filter $b^n[\boldsymbol{k}]=\phi(\k), \boldsymbol{k} \in \mbb{Z}^2$. Using \eqref{factor}, the desired projections can then be efficiently realized in two stages:	\\ 
(1) \textbf{Non-Adaptive Step}: The entire image $f$ is filtered once using the fixed-scale RS filter to give the pre-integrated image:
\begin{align}
\label{NA}
g_{\b}[\m]=\begin{cases} \Delta^{-N}_{\b} c[\m] ,& \text{with interpolation}  \\
 (u^N_{\b} \ast c)[\m],  &\text{without interpolation}
\end{cases}
\end{align}
This also gives us the continuous pre-integrated image $F(\x)=\sum_{\k \in \mbb{Z}^2} g_{\b}[\k]  (\phi \ast \bbeta_{\b}^N)(\x+\btau-\k)$;\\ \\
(2) \textbf{Adaptive Step}:  The pre-integrated image $g_{\b}$ is then locally filtered using the pointwise localization operator $\Delta^N_{\a(\m)}$ to give the adaptively filtered output
\begin{align}
\label{mask}
s[\m]=\Delta^N_{\a(\m)}F(\k)=\sum_{\k \in \mbb{Z}^2} w_{\a(\m)}[\k] g_{\b}[\m-\k]
\end{align}
at each location $\m \in \mbb{Z}^2$ corresponding to the scale-vector $\a(\m)$, where $w_{\a(\m)}[\k]=\Delta^N_{\a(\m)}\phi \ast \bbeta_{\b}^N(\k+\btau), \ \k \in \mbb{Z}^2,$ is the localization mask, and $\btau=0.5(\sum_{k=1}^N (a_k(\m)-b_k)\cos \theta_k, \sum_{k=1}^N (a_k(\m)-b_k)\sin \theta_k)$ is the pointwise shift-vector.
\begin{figure}
\centering
\includegraphics[width=0.7\linewidth]{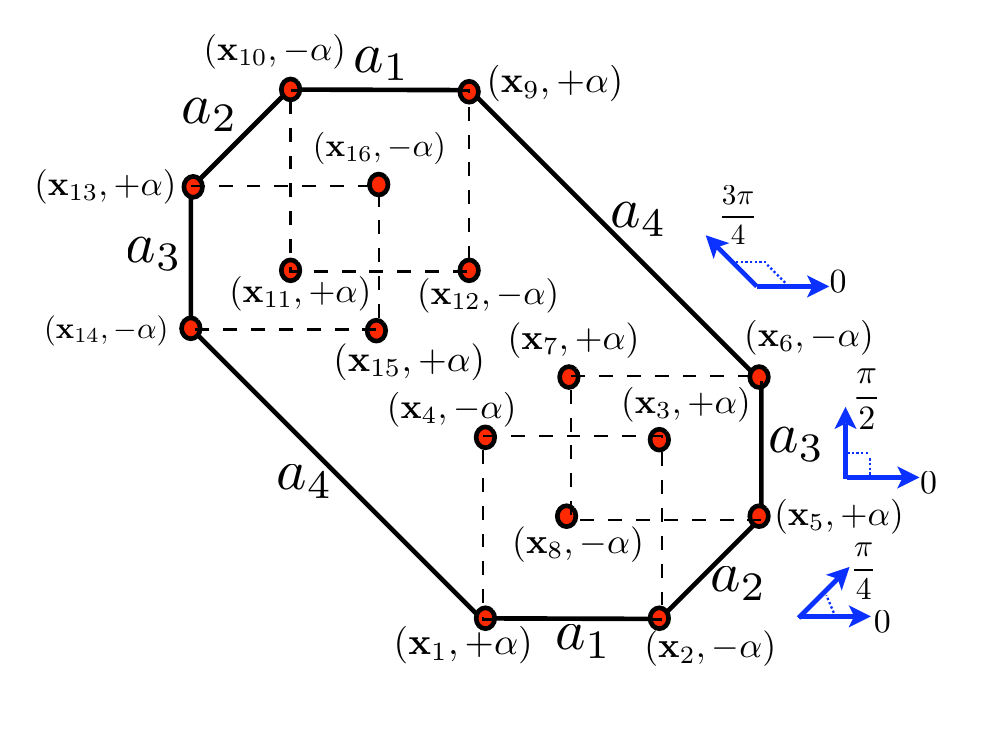}  
\caption{\bf Affine Mesh Geometry: The pair $(\cdot,\cdot)$ denotes the position of a mesh vertex and the corresponding weight.}
\label{mask_figure}
\end{figure} 

\section{Fast Elliptical Filtering Algorithm}

 Finally, we focus on the family of four-directional box spline $\bbeta^{4}_{\boldsymbol{a}}(\x)=(\varphi_{a_1,0} \ast \varphi_{a_2,\pi/4} \ast \varphi_{a_3,\pi/2} \ast \varphi_{a_4,3\pi/4})(\x),$ corresponding to the directional order $N=4$ and scale-vector $\a=(a_1,\ldots,a_4)$ in \eqref{def}. It is interesting to note that $2 \bbeta^{4}_{\a}(\x)$, corresponding to $\a=(1,\surd 2,1,\surd 2),$ is also known as the ZP (Zwart-Powell) element \cite{Zwart} in box spline literature. 
 
 Below, we outline the corresponding implementation aspects:\\ 
 \textbf{(1) Non-Adaptive Step}:  First, we simplify the pre-filtering by selecting the kernel $\phi(\x)=\delta(\x)$, with the result that $c[\k]=f[\k]$. Further, the fact that \eqref{NA} can be implemented (without interpolation) using the RS filter $u_{\b}^4[\k]$, with $\b=(1,\sqrt{2},1,\sqrt{2})$, further simplifies the computation. In particular, the pre-integrated image can be expressed as
\begin{align}
\label{RS}
g_{\b}[\k]=(u_{1,0} \ast u_{\sqrt{2},\pi/4} \ast u_{1,\pi/2} \ast u_{\sqrt{2},3\pi/4} \ast f)[\k]
\end{align}
Now, due to this tensor structure, \eqref{RS} can be then efficiently implemented in a recursive fashion in four steps, namely
\begin{description}
\item  (1) Computation of $F_0=u_{1,0} \ast f$ using $F_0[k_1,k_2]=f[k_1,k_2]+F_0[k_1-1,k_2]$.
\item (2) Computation of $F_{\pi/4}=u_{\sqrt{2},\pi/4} \ast F_0$ using $F_{\pi/4}[k_1,k_2]=\sqrt{2} F_0[k_1,k_2]+F_{\pi/4}[k_1-1,k_2-1]$.
\item (3) Computation of $F_{\pi/2}=u_{1,\pi/2} \ast F_{\pi/4}$ using $F_{\pi/2}[k_1,k_2]=F_{\pi/4}[k_1,k_2]+F_{\pi/2}[k_1,k_2-1]$. 
\item (4) Computation of $g_{\b}=u_{\sqrt{2},3\pi/4} \ast F_{\pi/2}$ as $g_{\b}[k_1,k_2]=\sqrt{2} F_{\pi/2}[k_1,k_2]+g_{\b}[k_1+1,k_2-1];$\\ 
\end{description}

\textbf{(2) Adaptive Step}: Finally, simplifying \eqref{mask}, we get 
\begin{align}
\label{final_eqn}
s[\m]=\sum_{i=1}^{16} \mathfrak{h}[i] g_{\interp}(\m+\btau-\x_i)
\end{align}
where $g_{\interp}(\x)=\sum_{\k} g_{\b}[\k] \bbeta^{4}_{\b}(\x-\k),$ is the ZP interpolation \cite{Zwart} of the discrete sequence $g_{\b}[\k],$ and $\mathfrak{h}[i]=(-1)^{i+1}\alpha, 1\leq i\leq 16$, are the non-zero weights of the affine FD mesh, with $\alpha=(a_1a_2a_3a_4)^{-1}.$ 

The corresponding (relative) positions $\{\x_i\}_{i=1}^{16} \subset \mbb{R}^2$ of the mesh vertices are shown in Table \ref{mask_positions}, with the convention $a'_j=a_j/\sqrt{2}$, for $j=2,4$. Figure \ref{mask_figure} gives the spatial ordering of the mesh vertices. The shift-vector in \eqref{final_eqn} is specified as $\btau=(\tau_1,\tau_2)$, where $\tau_1=(\sqrt{2}a_1+a_2-a_4-\sqrt{2})/2\sqrt{2}$ and $\tau_2=(a_2+\sqrt{2}a_3+a_4-3\sqrt{2})/2\sqrt{2}$.  Note that $\btau, \mathfrak{h}[i]$ and $\x_i$ are in fact defined pointwise in \eqref{final_eqn} using the scale-vector map $\a: \mbb{Z}^2 \rightarrow \mbb{R}^4_+$ (cf. \eqref{mask}); we dropped the pointwise index $\m$ just to simplify the equation.  
\begin{table}[htdp]
\bf\caption{Mesh Vertices}
\begin{center}
\begin{tabular}{|c|c|}
\hline
$\x_1:(0,0)$ &  $\x_9:(a_1+a'_2-a'_4,a_3+a'_2+a'_4)$ \\ \hline
$\x_2:(a_1,0)$ & $\x_{10}:(a'_2-a'_4,a_3+a'_2+a'_4)$ \\ \hline
$\x_3:(a_1,a_3)$ & $\x_{11}:(a'_2-a'_4,a'_2+a'_4)$ \\ \hline
$\x_4:(0,a_3)$ & $\x_{12}:(a_1+a'_2-a'_4,a'_2+a'_4)$ \\ \hline
$\x_5:(a_1+a'_2,a'_2)$ & $\x_{13}:(-a'_4,a_3+a'_4)$ \\ \hline
$\x_6:(a_1+a'_2,a_3+a'_2)$ & $\x_{14}:(-a'_4,a'_4)$ \\ \hline
$\x_7:(a'_2, a_3+a'_2)$ & $\x_{15}:(a_1-a'_4,a'_4)$ \\ \hline
$\x_8:(a'_2,a'_2)$ &  $\x_{16}:(a_1-a'_4,a_3+a'_4)$ \\ \hline
\end{tabular}
\end{center}
\label{mask_positions}
\end{table}

	Note that the algorithm has a fixed computational cost per output pixel as the size of the support of the FD mask in \eqref{final_eqn} is independent of the scale-vector. Specifically, the number of non-null weights is $4\times 4=16,$ i.e., $4$ clusters of $4$ points each, as shown in Figure \ref{mask_figure}.	

\section{Conclusion}
	We presented a novel elliptical filtering algorithm with fixed computational cost per output pixel. Our main goal was to formalize the adaptive elliptical filtering strategy using box splines and propose a fast and practical implementation algorithm. Computation of the scale-vector map presents a separate challenge in itself and will be discussed elsewhere.

\bibliographystyle{amsplain}
\bibliography{refs_adaptive.bib}

\end{document}